\documentclass[pra,twocolumn,floatfix,superscriptaddress]{revtex4-1}
\usepackage[utf8]{inputenc}
\usepackage{xcolor}
\usepackage[backref=none]{hyperref}
\hypersetup{
    colorlinks=true,          % false: boxed links; true: colored links
    linkcolor=blue,           % color of internal links
    citecolor=blue,           % color of links to bibliography
    urlcolor=olive,           % color of external links
}

\usepackage{amsmath}
\usepackage{amssymb}
\usepackage{multirow}
\usepackage{graphicx}
\graphicspath{ {images/} }
\usepackage{array}
\usepackage{upgreek}
\usepackage{mathtools}
\usepackage{cuted}
\usepackage[normalem]{ulem}

\usepackage[switch]{lineno}

\usepackage{comment}

\makeatletter

\begin{document}

%%%%% comment the following line
%\maketitle

%%%%% uncomment the following
\title{Efimov resonance position near a narrow Feshbach resonance in $^6$Li-$^{133}$Cs mixture}

\author{Ang Li}
\author{Yaakov Yudkin}
\affiliation{Department of Physics, QUEST Center and Institute of Nanotechnology and Advanced Materials, Bar-Ilan University, Ramat-Gan 5290002, Israel}

\author{Paul S. Julienne}
\affiliation{Joint Quantum Institute (JQI), University of Maryland and NIST, College Park, Maryland 20742, USA}

\author{Lev Khaykovich}
\affiliation{Department of Physics, QUEST Center and Institute of Nanotechnology and Advanced Materials, Bar-Ilan University, Ramat-Gan 5290002, Israel}

\date{\today}
%%%%% uncomment up to here

\begin{abstract}
In the vicinity of a narrow Feshbach resonances Efimov features are expected to be characterized by the resonance's properties rather than the van der Waals length of the interatomic potential.  
Although this theoretical prediction is well-established by now, it still lacks experimental confirmation. 
Here, we apply our recently developed three-channel model~\cite{Yudkin21} to the experimental result obtained in a mass-imbalanced $^6$Li-$^{133}$Cs mixture in the vicinity of the narrowest resonance explored to date~\cite{Johansen17}.
We confirm that the observed position of the Efimov resonance is dictated mainly by the resonance physics while the influence of the van der Waals tail of the interatomic potential is minor.
We show that the resonance position is strongly influenced by the presence of another Feshbach resonance which significantly alters the effective background scattering length at the narrow resonance position.
\end{abstract}

%%%%% uncomment the following line
\maketitle

\section{Introduction}
The Efimov effect in ultracold atoms emerges when the scattering length $a$ greatly exceeds the van der Waals length $r_{\text{vdW}}$ of the interatomic potential~\cite{Braaten&Hammer06}. 
The resonantly enhanced two-body interactions give rise to an infinite ladder of three-body bound states separated by a universal scaling factor.
Thus, to fully determine the three-body spectrum it suffices to do so for a single state.
Moreover, as the state's dependence on $a$ is described by a universal function, a single parameter is enough to define the entire spectrum.
For this matter it is convenient to choose the scattering length vale $a_{-}$ at which the ground state of the Efimov state meets the free-atom continuum.
Experimentally, this is the best studied parameter up to date~\cite{Greene17,Naidon17,D'Incao18}.
 
It was predicted that $a_{-}$ depends on the underlying two-body collisional resonance strength which is conveniently characterized by a dimensionless parameter $s_{res}$.
A collisional Feshbach resonance occurs when the free atoms in an open channel are coupled to a nearly degenerate two-body bound state in a closed channel~\cite{Chin10}:
For strong coupling with $s_{res}\gg 1$ (also known as the broad resonance regime) $a_{-}$ is universally related to $r_{\text{vdW}}$~\cite{Greene17,Naidon17}.
When the coupling weakens, $a_{-}$ deviates from this universality and, instead, tends to be dictated by the effective range of the Feshbach resonance for $s_{res}\ll 1$.
The latter regime can be described by a simplified theory with a square well potential tuned to have the same effective range as the true interaction potential~\cite{Petrov04}.

Experimental studies of narrow resonances are difficult due to the extreme magnetic field stability requirement.
The difficulties are two-fold.
First, the position of the Efimov resonance is predicted to be pushed towards higher scattering length values as compared to broad resonances which follow the Efimov-van der Waals universality. 
Second, narrow resonances are usually literally narrow, i.e. they are narrow functions of the magnetic field, which causes large changes in the scattering length over tiny variations of the magnetic field.
The combination of these two factors renders into an unrealistically tough requirement on the magnetic field stability, and hence this demanding regime was rarely approached experimentally~\cite{Roy13,Chapurin19}.
The narrowest resonance studied up to date is in the $^6$Li-$^{133}$Cs mixture~\cite{Johansen17}.

The few-body aspects of heteronuclear mixtures attracted significant interest in the last decade, both theoretical~\cite{DIncao06,Helfrich10,Petrov15,Acharya16,DIncao17,Giannakeas18,Rosa18,Sandoval18,Zhao19,Binh21,Giannakeas21} and experimental~\cite{Bloom13,Maier15,Wacker16,Kato17,Tung14,Pires14,Ulmanis16,Haefner17}.
The $^6$Li-$^{133}$Cs mixture is the most extreme mass imbalanced system in which Efimov features were observed up to date making it favorable for the attempt to reveal the few-body physics at a narrow Feshbach resonance.
In contrast to homonuclear systems, where the large universal scaling factor makes the observations of two consecutive Efimov resonances challenging~\cite{Huang14}, the large mass ratio in the Efimov favorable heavy-heavy-light scenario was predicted to decrease the scaling factor significantly~\cite{DIncao06}.
The Efimov physics in the $^6$Li-$^{133}$Cs mixture has been subject of intense experimental investigation in the vicinity of two broad Feshbach resonances and the decreased scaling factor was confirmed~\cite{Tung14,Pires14,Ulmanis16,Haefner17}. 
This motivated the attempt to look for Efimov features in the vicinity of a narrow Feshbach resonance despite the fact that no theoretical prediction is available in this region~\cite{Johansen17}.

Indeed, the position of the Efimov resonance was revealed at a larger scattering length as compared to the position predicted by the Efimov-van der Waals universality and measured in the vicinity of broad resonances~\cite{Johansen17}.
This result remains theoretically unexplored although developing a suitable theory can clarify several interesting aspects of the three-body physics at a narrow resonance.
For example:
How important is the van der Waals tail of the real interatomic potential compared to the resonance physics?
And:
What is the influence of a nearby overlapping Feshbach resonance?

Here we consider these questions by extending our recently developed three-channel theory to mass-imbalanced mixtures and applying it to the experimentally relevant resonances in the $^6$Li-$^{133}$Cs mixture.
We show that the position of the Efimov resonance is well-captured by this theory if the overlapping Feshbach resonances are properly taken into account.
To the best of our knowledge this is the first time such a theory demonstrates predictive power for Efimov physics in a real atomic system.
Based on this result we can place the upper bound for the contribution of the finite range of the interatomic potential (i.e. the van der Waals length) to the position of the Efimov resonance.
Unfortunately, the lack of other experimental results under similar conditions prohibits further bench-marking of our model.

\section{The model Hamiltonian}

Inspired by the two-channel model~\cite{Petrov04,Castin06,Gogolin08}, we develop a suitable model step-by-step, starting from an open channel of free atoms.
By considering a non-interacting open channel (zero background scattering) the short-range physics is neglected.
The resonant two-body interactions are modeled by coupling the open channel to a closed molecular channel which is detuned by a magnetic field-dependent binding energy.
The weakly coupled limit (narrow resonance) leads to a large effective range $r_e$ which significantly exceeds $r_{\text{vdW}}$~\cite{Gogolin08}.
More resonances can be included by coupling the open channel to additional closed channels~\cite{Yudkin21}.

We consider a $^6$Li-$^{133}$Cs mixture where both atoms are prepared in their respective absolute ground states ($aa$-channel).
At $893$~G there is a narrow Feshbach resonance which, according to coupled channels calculations using the model of Ref.~\cite{Tung13}, features a large and negative effective range at the resonance's position ($r_{e}=-1541 a_{0}$, where $a_{0}$ is the Bohr radius). 
As the van der Waals length of the Li-Cs interaction potential is $r_{\text{vdW}}= 44.8 a_{0}$, the narrow resonance criterion is well satisfied: $|r_{e}|\gg r_{\text{vdW}}$ or, alternatively, $s_{res}=0.0509\ll 1$.
Moreover, $a_{bg}=-30 a_{0}$ justifies the assumption of negligible background scattering~\cite{Repp13,Tung13,Ulmanis_2015}.
However, another Feshbach resonance at $843$~G is expected to play an important role.
This resonance is of intermediate character, being neither broad nor narrow.
As is shown below, it overlaps with the narrow resonance and strongly alters the local background scattering in the latter's vicinity.
Taking into account the $843$~G resonance is essential to reveal the predictive power of our three-channel model.

We start with the most generic case of three distinguishable atomic species (labeled $i=1,2,3$) with masses $m_{i}$.
Each atom pair can form a molecule in either of two closed channels $\nu=1,2$.
We define creation operators of atoms: $\hat{a}_{\vec{q},i}^{\dagger}$, and of molecules: $\hat{b}_{\vec{q},i,\nu}^{\dagger}$, where $\vec{q}$ denotes the particles momentum.
The index $i$ in $\hat{b}_{\vec{q},i,\nu}^{\dagger}$ labels the atom {\it not} part of the molecule.
The operators satisfy standard commutation relations.
The conversion of two atoms $i\neq j$ to a molecule $k\neq i,j$ in channel $\nu$ is most generally described by the term
\begin{equation}
\delta\left(\vec{q}_{1}-\vec{q}_{2}-\vec{q}_{3}\right)\hat{b}_{\vec{q}_{1},k,\nu}^{\dagger}\hat{a}_{\vec{q}_{2},i}\hat{a}_{\vec{q}_{3},j},
\end{equation}
where the $\delta(\vec{q}_{1}-\vec{q}_{2}-\vec{q}_{3})$ signifies momentum conservation.
%}

The total Hamiltonian consists of a bare atomic, a bare molecular and an interaction term:
\begin{equation}
\hat{H}=\hat{H}^{\left(\text{at}\right)}+\hat{H}^{\left(\text{mol}\right)}+\hat{H}^{\left(\text{int}\right)}.
\end{equation}
The bare atomic term is made of three parts, one for each species:
\begin{subequations}
\begin{equation}
\hat{H}^{\left(\text{at}\right)}=\sum_{i=1}^{3}\hat{H}_{i}^{\left(\text{at}\right)}
\end{equation}
\begin{equation}
\hat{H}_{i}^{\left(\text{at}\right)}=\int\frac{d^{3}q}{\left(2\pi\right)^{3}}\frac{\hbar^{2}q^{2}}{2m_{i}}\hat{a}_{\vec{q},i}^{\dagger}\hat{a}_{\vec{q},i}.
\end{equation}
The bare molecular term is made of six parts, one for each pair ($i$) and each channel ($\nu$):
\begin{equation}
\hat{H}^{\left(\text{mol}\right)}=\sum_{i=1}^{3}\sum_{\nu=1}^{2}\hat{H}_{i,\nu}^{\left(\text{mol}\right)}
\end{equation}
\begin{equation}
\hat{H}_{i,\nu}^{\left(\text{mol}\right)}=\int\frac{d^{3}q}{\left(2\pi\right)^{3}}\left(\frac{\hbar^{2}q^{2}}{2M_{i}}+E_{i,\nu}\right)\hat{b}_{\vec{q},i,\nu}^{\dagger}\hat{b}_{\vec{q},i,\nu},
\end{equation}
where the mass of a molecule is $M_{i}=(m_{j}+m_{k})$ and the energy detuning from the open channel is $E_{i,\nu}=\mu_{i,\nu}(B_{i,\nu}-B)$ with $\mu_{i,\nu}$ the differential magnetic moment and $B_{i,\nu}$ the bare resonance position.
Finally, the interaction term also consists of six parts:
\begin{equation}
\hat{H}^{\left(\text{int}\right)}=\sum_{k=1}^{3}\sum_{\nu=1}^{2}\hat{H}_{k,\nu}^{\left(\text{int}\right)}
\end{equation}
\begin{widetext}
\begin{equation}
\hat{H}_{k,\nu}^{\left(\text{int}\right)}=\frac{\Lambda_{k,\nu}}{2}\sum_{i,j\neq k}\int\frac{d^{3}q_{1}}{\left(2\pi\right)^{3}}\int\frac{d^{3}q_{2}}{\left(2\pi\right)^{3}}\left[\hat{b}_{\vec{q}_{1},k,\nu}^{\dagger}\hat{a}_{\vec{q}_{2}+\frac{\vec{q}_{1}}{2},j}\hat{a}_{-\vec{q}_{2}+\frac{\vec{q}_{1}}{2},i}+\hat{a}_{-\vec{q}_{2}+\frac{\vec{q}_{1}}{2},i}^{\dagger}\hat{a}_{\vec{q}_{2}+\frac{\vec{q}_{1}}{2},j}^{\dagger}\hat{b}_{\vec{q}_{1},k,\nu}\right],
\end{equation}
\end{widetext}
\end{subequations}
where the factor of $1/2$ avoids double-counting.
Note that we assume zero direct coupling between the two closed channels $\nu=1$ and $\nu=2$.
Without loss of generality, this coupling can be diagonalized by introducing a dressed basis in which interactions are absorbed by the energy shifts. 
A more rigorous approach considered in Ref.~\cite{Yudkin21} shows that this coupling adds an additional free parameter to the system which remains redundant when the other parameters are fixed by the two-body observables.
%Therefore, assuming no such direct coupling changes nothing in our analysis.
Indirect coupling through the common continuum remains intact.

\section{Three distinguishable particles}

\subsection{Two-body sector}

Since there are three distinct atomic species there are three two-body sectors $k=1,2,3$.
However, all three are permutations of each other.
The $k$-th two-body sector is descried by the Scr\"odinger equation $(\hat{H}-E)|\psi^{\left(\text{2B}\right)}_{k}\rangle=0$ and the (center-of-mass frame) two-body Ansatz is: 
\begin{equation}
|\psi^{\left(\text{2B}\right)}_{k}\rangle=\left(\sum_{\nu}\beta_{k,\nu}\hat{b}_{\vec{q}=0,k,\nu}^{\dagger}+\int\frac{d^{3}q}{\left(2\pi\right)^{3}}\alpha_{k}\left(\vec{q}\right)\hat{a}_{\vec{q},i}^{\dagger}\hat{a}_{-\vec{q},j}^{\dagger}\right)|0\rangle,
\label{eq:3-channel:two-body wave function}
\end{equation}
where $i\neq j\neq k\neq i$.
Scattering properties, in particular the scattering length $a_k$ and the effective range $r_{e,k}$, are derived from the positive energy solution $E=\hbar^2q_{k}^2/2 \mu_k>0$, while for $E=-\hbar^2\left(\lambda^{D}_{k}\right)^2/2 \mu_k<0$ the dimer binding energy is found. 
Here, $\mu_k=m_i m_j/(m_i+m_j)$ is the reduced mass of pair $i\neq j$.
Note that, for the sake of compact notation, the relative momentum $q_{k}$ of the free atoms can be formally related to the binding wave number $\lambda^{D}_{k}$ via $q_{k}=i\lambda^{D}_{k}$.

The two-body Scr\"odinger equation leads to the following two coupled equations ($\nu=1,2$):
\begin{multline}
%\begin{equation}
\tilde{\beta}_{k,\nu}\left(\tilde{E}_{k,\nu}-\tilde{q}_{k}^{2}\right)+\tilde{\Lambda}_{k,\nu}\Theta\left(E\right)\\-\frac{\tilde{\Lambda}_{k,\nu}}{2\pi^{2}}\left(1+\frac{i\pi}{2}\tilde{q}_{k}\right)
\sum_{\nu^\prime}\tilde{\Lambda}_{k,\nu^\prime}\tilde{\beta}_{k,\nu^\prime}
= 0,
\label{eq:3-channel:equations for 2-body sector (normalized)}
%\end{equation}
\end{multline}
where $\Theta\left(E\right)$ is the Heaviside step function.
In Eq.~(\ref{eq:3-channel:equations for 2-body sector (normalized)}) all quantities are renormalized with respect to the naturally occurring momentum cut-off $q_{c}$ and its associated energy $E_{c,k}=\hbar^2q_c^2/2\mu_k$ (see section~\ref{sec:gen 3body sector}).
A dimensionful quantity $x$ is denoted $\tilde{x}$ when normalized.

Solving Eq.~(\ref{eq:3-channel:equations for 2-body sector (normalized)}) for $E>0$ allows for computation of the scattering amplitude:
\begin{equation}
\tilde{f}\left(q_{k}\right)=-\sum_{\nu}\frac{\tilde{\Lambda}_{k,\nu}\tilde{\beta}_{k,\nu}}{4\pi}.
\label{eq:3-channel:equation for scattering amplitude (normalized)}
\end{equation}
The resulting expression is expanded to second order in $\tilde{q}_{k}$ and compared to the effective range expansion: $\tilde{f}^{-1}\left(q_{k}\right)=-\tilde{a}_k^{-1}-i\tilde{q}_{k}+\tilde{r}_{e,k}\tilde{q}_{k}^2/2$, to find the interspecies scattering length $\tilde{a}_k$ and the effective range $\tilde{r}_{e,k}$.
When $\tilde{q}_{k} = 0$ the solution of Eq.~(\ref{eq:3-channel:equations for 2-body sector (normalized)}) leads to an expression of the scattering length which can be directly compared to coupled-channel calculations.

For negative dimer energy $E<0$, Eq.~(\ref{eq:3-channel:equations for 2-body sector (normalized)}) leads to  a fourth-order polynomial equation for $\lambda^{D}_{k}$, whose positive roots correspond to the physically relevant solutions~\cite{Yudkin21}.

\subsection{Three-body sector} 
\label{sec:gen 3body sector}

The trimer binding energy $E_T=-\hbar^2\lambda_T^2/2\mu_T$, with $\lambda_T>\max(0,\lambda^{D}_{k})$, is the eigenvalue associated with the three-body wave function:
%\begin{widetext}
\begin{multline}
|\psi_{3B}\rangle=\sum_{i,\nu}\int\frac{d^{3}q}{\left(2\pi\right)^{3}}\beta_{i,\nu}\left(\vec{q}\right)\hat{b}_{\vec{q},i,\nu}^{\dagger}\hat{a}_{-\vec{q},i}^{\dagger}|0\rangle \\
+\int\frac{d^{3}q_1}{\left(2\pi\right)^{3}}\int\frac{d^{3}q_2}{\left(2\pi\right)^{3}}\alpha\left(\vec{q}_1,\vec{q}_2\right)\hat{a}_{-\vec{q}_2+\frac{\vec{q}_1}{2},1}^{\dagger}\hat{a}_{\vec{q}_2+\frac{\vec{q}_1}{2},2}^{\dagger}\hat{a}_{-\vec{q}_1,3}^{\dagger}|0\rangle.
\label{eq:3-channel:three-body wave function}
\end{multline}
%\end{widetext}
Direct substitution of $|\psi_{3B}\rangle$ into $(\hat{H}-E_T)|\psi_{3B}\rangle=0$ leads to seven coupled integral equations.
The first one, from projecting onto the free atom continuum, is
\begin{widetext}
\begin{subequations}
\begin{multline}
\alpha\left(\vec{q_{1}},\vec{q_{2}}\right)\left(\frac{\hbar^{2}\left|\vec{q_{2}}-\frac{\vec{q_{1}}}{2}\right|^{2}}{2m_{1}}+\frac{\hbar^{2}\left|\vec{q_{2}}+\frac{\vec{q_{1}}}{2}\right|^{2}}{2m_{2}}+\frac{\hbar^{2}q_{1}^{2}}{2m_{3}}-E_T\right) \\
+\sum_{\nu}\left[\Lambda_{1,\nu}\beta_{1,\nu}\left(\vec{q_{2}}-\frac{\vec{q_{1}}}{2}\right)+\Lambda_{2,\nu}\beta_{2,\nu}\left(-\vec{q_{2}}-\frac{\vec{q_{1}}}{2}\right)+\Lambda_{3,\nu}\beta_{3,\nu}\left(\vec{q_{1}}\right)\right]=0.
\end{multline}
The remaining six are structured as three pairs $\nu=1,2$:
\begin{equation}
\beta_{1,\nu}\left(\vec{q_{1}}\right)\left(\frac{\hbar^{2}q_{1}^{2}}{2\mu_{1}'}+E_{1,\nu}-E_T\right)+\Lambda_{1,\nu}\int\frac{d^{3}q_{2}}{\left(2\pi\right)^{3}}\alpha\left(\vec{q_{2}}-\frac{\vec{q_{1}}}{2},\frac{\vec{q_{2}}}{2}+\frac{3\vec{q_{1}}}{4}\right)=0
\end{equation}
\begin{equation}
\beta_{2,\nu}\left(\vec{q_{1}}\right)\left(\frac{\hbar^{2}q_{1}^{2}}{2\mu_{2}'}+E_{2,\nu}-E_T\right)+\Lambda_{2,\nu}\int\frac{d^{3}q_{2}}{\left(2\pi\right)^{3}}\alpha\left(-\vec{q_{2}}-\frac{\vec{q_{1}}}{2},\frac{\vec{q_{2}}}{2}-\frac{3\vec{q_{1}}}{4}\right)=0
\end{equation}
\begin{equation}
\beta_{3,\nu}\left(\vec{q_{1}}\right)\left(\frac{\hbar^{2}q_{1}^{2}}{2\mu_{3}'}+E_{3,\nu}-E_T\right)+\Lambda_{3,\nu}\int\frac{d^{3}q_{2}}{\left(2\pi\right)^{3}}\alpha\left(\vec{q_{1}},-\vec{q_{2}}\right)=0
\end{equation}
\label{eq:3-channel:coupled equations}
\end{subequations}
\end{widetext}
where $\mu^\prime_k = M_i m_i/(M_i+m_i)$ is the reduced mass of the molecule and the free atom.

We note that these equations reduce to the previously derived homo-nuclear three-channel model for $i=j=k$ and to the hetero-nuclear two-channel model in the case $\Lambda_{i,2}=0$.
To proceed, the free particle amplitude $\alpha\left(\vec{q_{1}},\vec{q_{2}}\right)$ is eliminated from the first equation and plugged into the others.
The first of the two integrals in each equation can be solved, as in the two-body sector, by introducing a high momentum cut-off $q_c$ with which the coupling constants are renormalized according to $\tilde{\Lambda}_{k,\nu}=\Lambda_{k,\nu} q_{c}^{3/2}/E_{c}$, and the amplitudes according to $\tilde{\beta}_{k,\nu}=\beta_{k,\nu} q_{c}^{3/2}$.
The renormalized magnetic moment is $\tilde{\mu}_i=\mu_i/E_{c}$ and all momenta are $\tilde{q}=q/q_c$.
In addition one uses the $s$-wave property that $\beta_{k,\nu}(\vec{q})=\beta_{k,\nu}(q)$ are spherically symmetric.
One thus ends up with six one-dimensional coupled integral equations.

\section{Lithium-Cesium-Cesium system}

While Eqs.~(\ref{eq:3-channel:coupled equations}) are too complex for solve in general, they serve as a convenient starting point to study specific cases.
Here, we apply the model to the 2+1 case, i.e. two particles with equal masses and one distinguishable particle, of $^6$Li-$^{133}$Cs-$^{133}$Cs trimers.

\subsection{Two-body sector}

For the remainder of the paper we define the relevant masses: $m=m_{\text{Li}}$ and $M=m_{\text{Cs}}$.
In the two-body sector, only one interspecies molecule is possible (LiCs) such that the index $k$ can be omitted in Eq. (\ref{eq:3-channel:equations for 2-body sector (normalized)}).
Solutions of the remaining two equations for $E>0$ are compared to coupled channel calculations~\cite{Tung13} to fix the free parameters of the model.
Here we consider the $aa$ collisional channel of the $^{6}$Li-$^{133}$Cs mixture, where both atoms are polarized on their respective absolute ground states, and which is relevant for the experiment of Ref.~\cite{Johansen17}.

We proceed in the following way.
We fit the magnetic field dependence of the scattering length provided by coupled channel calculations with the well-known parametrization expression:
\begin{equation}
\tilde{a}_{\text{LiCs}}(B)=\frac{\tilde{\Delta}_1}{B^{(\text{res})}_{1}-B}+\frac{\tilde{\Delta}_{2}}{B^{(\text{res})}_{2}-B},
\label{eq:3-channel:Feshbach resonance formula}
\end{equation}
where the resonance widths $\tilde{\Delta}_{\nu}$ and positions $B^{(\text{res})}_{\nu}$ are experimental observables.
These observable parameters are conveniently related to the model's bare parameters via analytic expressions~\cite{Yudkin21} with which the latter are found (see Table~\ref{tb:three channel model parameters}).
The differential magnetic moments $\tilde{\mu}_{\nu}$ are not fitting parameters.
Instead, they are extracted from the asymptotic behavior of the coupled channel dimer binding energies.
In real units they are, $\mu_{1}=- h\times 3.03$~MHz/G and $\mu_{2}=-h\times 2.84$~MHz/G.

\begin{table}[b]
\centering
\begin{tabular}{c | c}
\hline\hline
$\Delta_1/a_0$ (G) & $1741.13$ \\
$\Delta_2/a_0$ (G) & $131.351$ \\
\hline
$B_1-B_2^{(\text{res})}$ (G) & $-68.736$ \\
$B_2-B_2^{(\text{res})}$ (G) & $-1.01$ \\
$\tilde{\Lambda}_1$ & $3.50$ \\
$\tilde{\Lambda}_2$ & $0.707$ \\
\hline\hline
\end{tabular}
\caption{\label{tb:three channel model parameters}
Parameters of the three channel model derived from fitting Eq.~(\ref{eq:3-channel:Feshbach resonance formula}) to coupled channel calculations of $^6$Li-$^{133}$Cs.}
\end{table}

\begin{figure}
\centering
\includegraphics[width=1\linewidth]{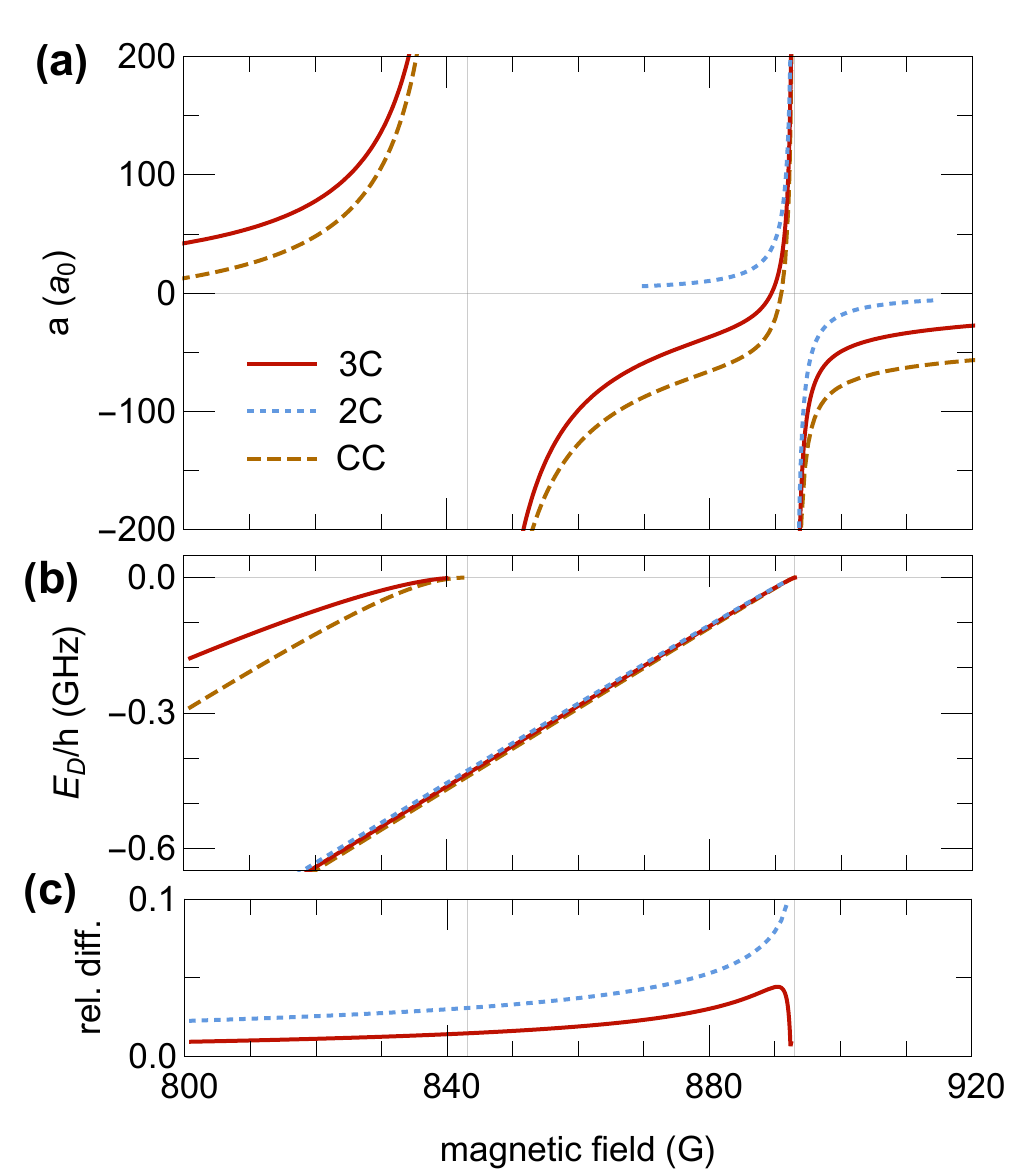}
\caption{\label{fig:LiCs 2-body sector}
Two-body sector of LiCs.
(a) Magnetic field dependence of the scattering length, (b) binding energy and (c) the relative difference of the binding energy.
In (a) and (b) the dashed brown line corresponds to the coupled channel calculations and red solid (blue dotted) line represents the three-(two-) channel model.
In (c) the red solid (blue dotted) curve shows the relative difference between the coupled-channels and the three-(two-)channel models.
The difference with the three-channel model is consistently lower at all magnetic field values than with the less successful two-channel model.}
\end{figure}

In Fig.~\ref{fig:LiCs 2-body sector}(a) the scattering length of the three-channel model as a function of the magnetic field, which by construction coincides with Eq.~(\ref{eq:3-channel:Feshbach resonance formula}), is compared to the coupled channel calculations together with the result of the two-channel model.
The agreement is very good in the vicinity of the Feshbach resonances. 
The discrepancies between the coupled channel calculations and the three-channel model are visible for small absolute values of the scattering length. 
This is because our model does not include the global background scattering length.
The two-channel model, also shown in the figure, is significantly less successful at capturing the coupled channel calculations.
Naturally, the model includes only one closed channel and  hence only a single Feshbach resonance.
The absence of the scattering length zero-crossing leads to a significant disagreement between the model and the coupled channel calculations.

In Fig.~\ref{fig:LiCs 2-body sector}(b) the binding energies of the dimers from the coupled channel calculations are compared to the results of the two- and three-channel model.
Both models are successful in describing the narrow resonance and capture the energy level down to hundreds of MHz.
However, a closer look at the differences between the models [shown in Fig.~\ref{fig:LiCs 2-body sector}(c)] emphasizes that the three-channel model is a more successful approach to the real system.
The three-channel model also reproduces the biding energy of the intermediate resonance although good agreement is obtained only in the regime of weak binding.
This discrepancy might be explained by the intermediate character of the underlying narrow Feshbach resonance for which our model model's assumptions cease to be valid.

In addition, we found the effective range $r_e=-1743 a_0$ (at resonance) to differ by 4 percent from the resonance contribution $-1666\, a_0$ to the effective range. 
The latter value is found by subtracting the van der Waals contribution $+125\, a_0$~\cite{Gao98} from the coupled channels value $-1541\, a_0$, which includes the sum of the van der Waals and resonant contributions~\cite{Gao11}.
%The latter is calculated to be $r_e=-1722 a_0$ according to the prescription of Ref.~\cite{Gao11}.

In conclusion, the two-body sector reveals that the three-channel model is a better way to describe the real Li-Cs interactions in the $aa$ collisional channel, due to the intermediate Feshbach resonance overlapping with the narrow one and affecting the latter's properties.
Adding a third channel is a necessary procedure.

Note that in this particular case there is an alternative theoretical approach.
The two-channel model can be extended to include a non-zero background scattering length~\cite{Werner09} which is expected to improve the agreement with the coupled channels calculations.
This approach has its own limitations partially discussed in Ref.~\cite{Yudkin21} and it has not yet been extended to the mass-imbalanced mixtures.
The three-channel model is superior because it takes the background scattering length into account by considering its real cause, namely the presence of another Feshbach resonance in close proximity.

\subsection{Three-body sector}
\label{sec:LiCsCs 3-body}

\begin{figure}
\centering
\includegraphics[width=1.\linewidth]{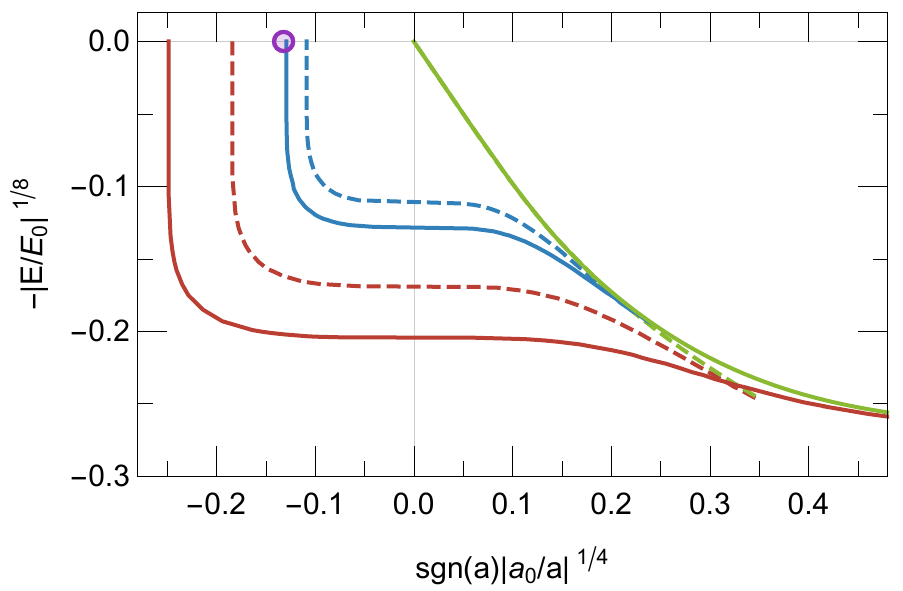}
\caption{\label{fig:three-body}
Three-body sector of LiCsCs.
The three-channel model (solid) is compared to the two-channel model (dashed).
Shown are the dimer (green) and the ground (red) and first excited (blue) Efimov states.
The purple data point is the measurement from Ref.~\cite{Johansen17}.}
\end{figure}

%For the LiCsCs three-body sector the original six equations described in Sec.~\ref{sec:gen 3body sector} reduce to four.
For the LiCsCs three-body sector, Eqs.~(\ref{eq:3-channel:coupled equations}) reduce to four coupled equations.
The four remaining molecular amplitudes $\beta_{i,\nu}$ are $i=\{$LiCs,CsCs$\}$ and $\nu=\{1,2\}$.

Further simplifications come from the fact that we neglect Cs-Cs interactions by setting the relevant scattering length ($a_{\text{CsCs}}$) to zero.
In reality, its value is moderate and positive in the vicinity of the narrow Feshbach resonance ($a_{\text{CsCs}}=260 a_{0}$), while it is large and negative at the intermediate resonance ($a_{\text{CsCs}}=-1400 a_{0}$)~\cite{Berninger13}.
Since we consider the Efimov spectrum in the close vicinity of the narrow resonance the latter value is irrelevant.
On the other hand, one should be aware of the positive $a_{\text{CsCs}}$, considering the fact that $a_{\text{CsCs}}>0$ affects the Efimov spectrum measured at intermediate Feshbach resonances~\cite{Ulmanis16,Haefner17,Johansen17}.
There, its main influence is to eliminate the ground state of the Efimov spectrum~\cite{Ulmanis16,Haefner17}.
Indeed, also in the vicinity of the narrow resonance, the ground Efimov state was not detected~\cite{Johansen17}.
Additionally, the first excited Efimov resonance in the vicinity of two intermediate resonances was measured to be within $\sim20\%$ from each other for both positive $a_{\text{CsCs}}=200a_{0}$ and large negative $a_{\text{CsCs}}=-1400a_{0}$ Cs-Cs scattering lengths~\cite{Ulmanis16,Johansen17}.
Thus, $\sim20\%$ can be considered the upper limit for our error if $a_{\text{CsCs}}$ is set to zero. 
Note, however, that at the narrow resonance, $a_{\text{CsCs}}$ is at least an order of magnitude smaller than the absolute values of the effective range and the interspecies scattering length at which the the first excited Efimov energy level crosses the threshold.
Therefore, its influence on the position of the Efimov resonance is expected to be less important than in the case of intermediate resonances.

This simplification leads to two coupled integral equations for $\beta_{LiCs,\nu}=\beta_\nu$:
\begin{widetext}
\begin{multline}
\left(\frac{\hbar^{2}q^{2}}{2\mu^\prime}+E_{\nu}-E_T\right)\beta_\nu\left(\vec{q}\right)-\frac{\mu\Lambda_{\nu}}{\pi^{2}\hbar^{2}}\left(q_{c}-\frac{\pi}{2}\sqrt{\frac{(2r+1)q^{2}+r(r+1)\frac{m}{\mu_{T}}\lambda_T^2}{(r+1)^{2}}}\right)\sum_{\nu^\prime}\Lambda_{\nu^\prime}\beta_{\nu^\prime}\left(\vec{q}\right) \\
-\frac{m\Lambda_{\nu}}{4\pi^{2}\hbar^{2}}\int_{0}^{\infty}dp\ln\left(\frac{p^{2}+\frac{2r}{r+1}pq+q^{2}+\frac{r}{r+1}\frac{m}{\mu_{T}}\lambda_T^{2}}{p^{2}-\frac{2r}{r+1}pq+q^{2}+\frac{r}{r+1}\frac{m}{\mu_{T}}\lambda_T^{2}}\right) \sum_{\nu^\prime}\Lambda_{\nu^\prime}\beta_{\nu^\prime}\left(\vec{p}\right)=0,
\label{eq:3-channel:two coupled equations in terms of k,q}
\end{multline}
%\end{widetext}
where $r=M/m$ is the mass ratio. 
Following the procedure shown in Ref.~\cite{Yudkin21}, we represent the two three-body scattering amplitudes as a vector: $\psi(q)=[\beta_1(q),\beta_2(q)]^T$, and the coefficients of Eqs.~(\ref{eq:3-channel:two coupled equations in terms of k,q}) as a $2\times2$ matrix: $\mathcal{M}_{\lambda_T}\left(q_1,q_2\right)$ that depends on $\lambda_T$.
Then Eqs.~(\ref{eq:3-channel:two coupled equations in terms of k,q}) take the form $\int_0^\infty dq_2 \mathcal{M}_{\lambda_T}(q_1,q_2)\psi(q_2)=0$ and a non-trivial solution is obtained for $\det\mathcal{M}_{\lambda_T}(q_1,q_2)=0$.
We perform renormalization as in Sec.~\ref{sec:gen 3body sector}, use the practical substitution:
\begin{equation}
\tilde{q}_i=\sqrt{\frac{r(r+1)}{(2r+1)}\frac{m}{\mu_{T}}}\tilde{\lambda}\sinh\xi,
\end{equation}
and Eqs.~(\ref{eq:3-channel:two coupled equations in terms of k,q}) become
\begin{equation}
\int_{-\infty}^{\infty} d\xi \mathcal{M}_{\lambda_T}\left(\xi,\xi^\prime\right)\psi\left(\xi^\prime\right)=0.
\label{eq:3-channel:int d xi M(xi,xi') psi(xi') = 0}
\end{equation}
Extension of the lower integration limit to $-\infty$ requires that both $\tilde{\beta}_1(\xi)$ and $\tilde{\beta}_2(\xi)$ be odd functions of $\xi$.
The vector $\psi(\xi)$ and the matrix elements are
\begin{subequations}
\begin{equation}
\psi(\xi)=\left[\tilde{\beta}_1(\xi),\tilde{\beta}_2(\xi)\right]^T
\end{equation}
\begin{equation}
\left(\mathcal{M}_{\lambda_T}\right)_{ij}=\left[f_i\left(\xi^\prime\right)\delta_{ij}-\tilde{\Lambda}_i\tilde{\Lambda}_j g\left(\xi^\prime\right)\right]\delta\left(\xi-\xi^\prime\right)-\tilde{\Lambda}_i\tilde{\Lambda}_j L\left(\xi,\xi^\prime\right),
\end{equation}
\end{subequations}
where
\begin{subequations}
\begin{equation}
f_{i}\left(\xi\right)=\tilde{\lambda}_T\cosh\xi+\frac{\tilde{\mu}_{i}}{\tilde{\lambda}_T\cosh\xi}\left(B_{i}-B\right),
\end{equation}
\begin{equation}
%g\left(\xi\right)=\frac{1}{\pi^{2}}\left(\frac{1}{\tilde{\lambda}_T\cosh\xi}-\frac{\pi}{2}\right),
g(\xi)=\frac{1}{2\pi^{2}}\frac{\mu}{\mu_{T}}\left(\frac{1}{\tilde{\lambda}\cosh\xi}-\frac{\pi}{2}\sqrt{\frac{r}{r+1}\frac{m}{\mu_{T}}}\right)
\end{equation}
%\begin{widetext}
\begin{equation}
L(\xi,\xi')=\frac{1}{16\pi^{2}}\frac{m}{\mu_{T}}\sqrt{\frac{r(r+1)}{(2r+1)}\frac{m}{\mu_{T}}}\ln\left(\frac{\sinh^{2}\xi'+\frac{2r}{r+1}\sinh\xi'\sinh\xi+\sinh^{2}\xi+\frac{(2r+1)}{(r+1)^{2}}}{\sinh^{2}\xi'-\frac{2r}{r+1}\sinh\xi'\sinh\xi+\sinh^{2}\xi+\frac{(2r+1)}{(r+1)^{2}}}\right).
\end{equation}
%\end{widetext}
\end{subequations}
\end{widetext}
The requirement of a vanishing determinant:
\begin{equation}
\det \mathcal{M}_{\lambda_T}\left(\xi,\xi^\prime\right)=0,
\label{eq:3-channel:equantion for lambda_T, determinant = 0}
\end{equation}
defines a closed equation for $\lambda_T$.
In general, there are many values $\lambda_T=\lambda_T^{(\text{sol})}$ for which Eq.~(\ref{eq:3-channel:equantion for lambda_T, determinant = 0}) is satisfied however not all of them correspond to physical solutions.
To identify the real three-body bound sates one must compute the zero-eigenvalue eigenfunction $\psi(\xi)$ of $\mathcal{M}_{\lambda_T^{(\text{sol})}}$ in accordance with Eq.~(\ref{eq:3-channel:int d xi M(xi,xi') psi(xi') = 0}) and determine $\tilde{\beta}_1(\xi)$ and $\tilde{\beta}_2(\xi)$.
Then, the mathematical solution $\lambda_T^{(\text{sol})}$ is physical only if both are odd functions of $\xi$.
In addition, the number of nodes in $\tilde{\beta}_1(\xi)$ and $\tilde{\beta}_2(\xi)$ allow assignment of $\lambda_T^{(\text{sol})}$ to the ground or an excited Efimov state (see Sec. IV in Ref.~\cite{Yudkin21} for details).

To solve Eq.~(\ref{eq:3-channel:equantion for lambda_T, determinant = 0}) numerically, each block $\mathcal{M}_{ij}$ is represented as a $n\times n$ matrix by discretizing $\xi$ and $\xi^\prime$ in the interval $\left[-\xi_m,\xi_m\right]$ and step size $d\xi=2\xi_m/(n-1)$.
The total matrix thus has dimensions $2n\times2n$ and its determinant is found.
The computed ground and first excited states are shown in Fig.~\ref{fig:three-body}, where we used $\xi_m=20.02$ and $n=200$ (and $n=1600$ for some points) together with the parameters of Table~\ref{tb:three channel model parameters}.

\section{Discussion and Conclusions}

In Table~\ref{tb:theory experiment}, a comparison between the position of the first excited Efimov resonance predicted by the two- and three-channel models and the experimental result from Ref.~\cite{Johansen17} is presented.
The two-channel model overestimates the position of the resonance by more than a factor of two.
In contrast, the three-channel model agrees quite well with the experimental value.
For comparison the universal theory prediction is also listed. 
The latter is based on a single-channel model of Refs.~\cite{Ulmanis16,Wang13} and presented in Ref.~\cite{Johansen17}.
It is important to emphasize the amazing and not at all obvious fact that the overlapping resonances worked in favor of the experimental observation of the Efimov resonance in this particular case.

%{\lk
%Note, that here we consider exclusively the first excited Efimov state.
%This is because the ground state does not exist in real system due to finite and positive Cs-Cs scattering length (as discussed in section~\ref{sec:LiCsCs 3-body}).
%Our model can not possibly capture this behavior even and the apparent presence of the ground state should simply be ignored.
%}

Note, that our comparison between theory and experiment is limited to the first excited Efimov state.
Our minimal model does not capture the absence of the ground state, caused by the finite and positive Cs-Cs scattering length (see discussion in Sec.~\ref{sec:LiCsCs 3-body}).
%{\lk \sout{Combining this disagreement with the agreement of the first excited state indicates that the Efimov states decouple from the Cs-Cs interactions in the range $a_-^{(1)}<a_{\text{LiCs}}<a_-^{(2)}$}}.

In summary, the results presented in this paper confirm that the Feshbach resonance used in the experiment is narrow enough to effectively decouple the three-body physics from the van der Waals universality.
The remaining influence of the van der Waals length can then be estimated to be about $10\%$.
This estimation, however, is within the limits of the above-mentioned conservative error set by the $a_{\text{CsCs}}=0$ assumption.
Therefore, the upper bound for the influence of the finite range of the interaction potential is dominated by the latter, and can thus be quoted as $\lesssim 20\%$.

\begin{table}[h]
\centering
\begin{tabular}{l | l}
Source & $a_- ^{(2)} (a_0)$\\
\hline\hline
Experiment~\cite{Johansen17} & $-3,330(240)$ \\
Three-channel theory & $-3,600$ \\
Two-channel theory & $-7,189$ \\
Universal theory & $-2,200$ \\
\hline\hline
\end{tabular}
\caption{\label{tb:theory experiment}
The experimental value of the resonance position is contrasted to the various theory values.
The universal theory result is cited as per Ref.~\cite{Johansen17}.
%The second Efimov resonance position from the experiment, two- and three-channel models analyzed here and from the Efimov van-der Waals universality.
}
\end{table}

\section*{Acknowledgments}

We acknowledge fruitful discussions with F. Chevy and J. P. D'Incao.
This research was supported in part by the Israel Science Foundation (Grant No. 1543/20) and by a grant from the United States-Israel Binational Science Foundation (BSF), Jerusalem, Israel, and the United States National Science Foundation (NSF).

%\bibliography{fewbody}% Produces the bibliography via BibTeX.
%\bibliographystyle{unsrt}

\end{document}